\documentclass[runningheads]{llncs}

\usepackage[T1]{fontenc}
\usepackage{graphicx}
\usepackage{cite}
\usepackage{algorithm}
\usepackage{algorithmic}
\usepackage{amsmath}
\usepackage{amssymb}
\usepackage[utf8]{inputenc}
\begin{document}

\title{AdaFedXplain: Adaptive Federated Learning with Explainable AI for Ransomware Detection}
\titlerunning{Federated Learning with XAI for Ransomware Detection}
\author{
Md. Emran Biswas\inst{1}\and
Md. Delowar Hossain\inst{2} \and
Mst. Tanbin Yasmin Tanny\inst{1} \and
Muhammad Aminur Rahaman\inst{3} \and
Md Al Amin\inst{4} \and
Tangina Sultana\inst{1}\textsuperscript{*}
}

\institute{Department of Electronics and Communication Engineering, \and Department of Computer Science and Engineering, Hajee Mohammad Danesh Science \& Technology University, Bangladesh \\ 
\and
Department of Data Science and Engineering, Bangladesh University of Business and Technology, Bangladesh \\
 \and
Department of Computer Science, Colorado State University, Fort Collins, USA \\
\emph{\textsuperscript{*}Corresponding email:} tangina@hstu.ac.bd
}

\authorrunning{Biswas et al.}

\maketitle

\begin{abstract}
Attacks using ransomware have developed into complex multi-phase campaigns that target vital infrastructure, resulting in annual damages of billions of dollars. Conventional detection techniques are ineffective against zero-day variations, and privacy issues are raised by centralized machine learning algorithms. Federated learning allows for cooperative model training without data sharing, however current FL approaches suffer from explainability issues, high communication cost, and non-IID data heterogeneity. In this study, we present AdaFedXplain, an explainable AI-based adaptive federated learning system for ransomware detection. Our approach includes progressive XGBoost complexity to avoid overfitting, a decaying learning rate schedule for stable convergence, and dynamic client grading based on ransomware prevalence and data sufficiency. Using 62,485 real-world ransomware samples from four OS-based non-IID clients, we compare AdaFedXplain against FedAvg, FedProx, and SCAFFOLD. According to experimental data, AdaFedXplain outperforms all baselines with 99.59\% detection accuracy and 99.64\% F1-score while cutting energy usage by 75\%, latency by 65\%, and communication overhead by 75\%. The ablation investigation demonstrates that every element enhances performance. Security analysts can use SHAP explainability to identify important PE header characteristics that influence detection decisions. Our system provides a collaborative ransomware detection solution that is efficient and interpretable, while reducing privacy risks through federated learning's data-local training paradigm.

\keywords{Federated Learning \and Ransomware Detection \and Explainable AI \and AdaFedXplain \and Non-IID Data \and Machine Learning, Cybersecurity}
\end{abstract}

\section{Introduction}

The scope and complexity of ransomware attacks have increased dramatically, making them one of the most serious cybersecurity risks in the world. Modern ransomware has developed from simple file encryption to complex multi-stage attacks that target government infrastructure, healthcare, and banking. According to recent research, the frequency of attacks has been steadily increasing, and the yearly global damages are expected to exceed trillions of dollars \cite{azubuike2024economic,gharbi2025iot}. The need for strong defenses is highlighted by the growing use of sophisticated evasion strategies.

Heuristic analysis, manual reverse engineering, and signature-based detection are the mainstays of conventional ransomware mitigation techniques. Although these techniques work well against known families, they are unable to identify polymorphic and zero-day assaults, produce a large number of false positives, and scale poorly \cite{alzahrani2025survey,gazzan2025deep}. By using automated feature learning and behavior-based detection, machine learning gets around these restrictions and allows for generalization to previously unknown attack patterns without the need for signatures \cite{shahzadi2025healthcare,chukwuani2025ml}.

In this research, we propose a federated learning architecture that allows collaborative model training across remote clients without revealing sensitive data, thereby addressing the shortcomings of centralized ML-based ransomware detection. Our solution specifically addresses the non-IID heterogeneity present in real-world deployments, where various businesses display differing operating system distributions and ransomware prevalence, in contrast to prior FL approaches that assume IID data distributions. On a real-world ransomware dataset of 62,485 samples, we do a thorough comparison of three FL settings, assessing XGBoost model across four non-IID clients divided by MajorOSVersion.

Our proposed AdaFedXplain framework addresses the non-IID challenge by partitioning training data across four heterogeneous clients based on MajorOSVersion (Windows NT/2000, XP/2003, Vista/7, and 10/11), reflecting real-world OS diversity. For each communication round, we compute a client score \(S(i) = 0.6 \cdot \rho_i + 0.4 \cdot \min(1, |D_i|/10000)\), where \(\rho_i\) is the ransomware prevalence ratio, and select the top 60\% of clients for aggregation. We employ a decaying learning rate \(\eta_t = \max(0.01, 0.1 \cdot (1 - t/T))\) to ensure stable convergence and progressive XGBoost complexity (\(E_t = 60 + t\) estimators) to prevent overfitting.  Experimental results demonstrate that AdaFedXplain achieves 99.59\% detection accuracy and 99.64\% F1-score using XGBoost, outperforming FedAvg, FedProx, and SCAFFOLD. Critically, our method reduces communication overhead by 75\% (80 MB → 20 MB), latency by 65\% (4.0s → 1.4s), and energy consumption by 75\% (40J → 10J). Per-client F1-scores consistently exceed 99.5\%, while SCAFFOLD exhibits performance degradation with per-client F1-scores dropping to approximately 98.3\%, confirming the robustness of AdaFedXplain across heterogeneous OS environments.

Our main contributions are:

\begin{itemize}
\item For ransomware detection in non-IID contexts, we propose AdaFedXplain, an adaptive federated learning framework with dynamic client selection and decaying learning rates.

\item We incorporate SHAP explainability for ransomware detection, offering force and summary charts for transparent interpretation of model choices.

\item We provide the first thorough evaluation of four FL settings (FedAvg, FedProx, SCAFFOLD, and AdaFedXplain) using four OS-based non-IID clients and real-world ransomware samples.

\item AdaFedXplain delivers substantial improvements in resource efficiency over baseline methods, significantly reducing energy consumption, communication overhead, and latency, while maintaining high detection accuracy.

\end{itemize}

This is how the rest of the paper is structured. Related research on explainable AI, ransomware detection, and federated learning is reviewed in Section 2. The suggested AdaFedXplain methodology, which includes client selection, adaptive learning, and aggregation, is described in full in Section 3. The dataset, evaluation measures, and experimental design are explained in Section 4. The experimental results are presented and discussed in Section 5. The paper is concluded and next work is outlined in Section 6.

\section{Literature Review}

Federated Learning (FL) has become a potential paradigm for cybersecurity applications that protect privacy, especially in malware categorization and intrusion detection systems (IDS). According to recent research, FL preserves data confidentiality in sensitive situations like IoT and cloud systems by enabling distributed model training across numerous clients without exchanging raw data \cite{bilal2025flxai,latif2025flids}. To manage distributed optimization, popular aggregation algorithms like FedAvg and its extension FedProx have been frequently used. However, because of varied network settings, device-specific behaviors, and different threat profiles, their performance drastically deteriorates under non-IID data distributions, which are common in real-world cybersecurity datasets. Even though FedProx adds a proximal term to stabilize local updates, previous research shows that dealing with severely skewed client data continues to present convergence, fairness, and robustness issues \cite{bilal2025flxai,latif2025flids}. These drawbacks highlight the urgent need for flexible and context-aware FL techniques designed especially for cybersecurity jobs like ransomware detection.

The capacity of machine learning-based ransomware detection to recognize intricate assault patterns that are beyond the scope of conventional signature-based techniques has drawn a lot of interest. Ransomware has been classified using both static data (file entropy, opcode sequences, API call frequency) and dynamic behaviors (system call traces, execution patterns) using techniques like XGBoost, Logistic Regression, and deep neural networks. According to recent research, deep learning techniques are excellent at identifying temporal and behavioral patterns in execution traces, whereas gradient boosting models like XGBoost achieve high detection accuracy because of their resilience to high-dimensional and heterogeneous data \cite{yadav2026fatex,liu2026ids}. Despite their efficacy, these models mostly function in centralized environments, which raises serious issues with data privacy, scalability, generalization across distributed organizational environments, and regulatory compliance (e.g., GDPR, HIPAA). The integration of FL with ransomware detection is motivated by this centralized constraint.

In security applications, where interpretability and transparency are crucial for regulatory compliance, forensic investigation, and analyst trust, Explainable Artificial Intelligence (XAI) has grown in importance. By measuring the contribution of each input feature to model predictions, techniques like SHAP (SHapley Additive exPlanations) give feature-level attribution, allowing security teams to comprehend why a certain activity is reported as dangerous. In order to improve interpretability and facilitate post-incident forensic analysis, recent research incorporates SHAP and other XAI approaches into malware and intrusion detection frameworks \cite{liu2026ids,arepalli2026fplstm}. Despite these developments, there is still a significant gap in explainable, privacy-preserving security solutions due to the extremely restricted integration of XAI into distributed and federated environments. Several research gaps still exist despite significant advancements. First, no previous effort has combined detection models specific to ransomware with adaptive federated learning. Second, even though OS-based heterogeneity is a crucial feature of real-world ransomware datasets where various operating systems display distinct susceptibility patterns, existing FL frameworks fail to sufficiently handle non-IID data distributions resulting from this phenomenon. Third, although XAI techniques have been used in centralized environments, there isn't a comprehensive explainable federated learning framework created especially for ransomware detection. These gaps inspire us to work hard.

To address these challenges, we propose \textit{AdaFedXplain}, an adaptive federated learning framework with integrated explainability for ransomware detection. We model non-IID data distribution by partitioning clients based on \texttt{MajorOSVersion}, creating four clients corresponding to OS versions $\{4,5,6,10\}$ (Windows NT/2000, XP/2003, Vista/7, and 10/11). A client scoring function is defined as $S(i) = 0.6 \cdot \rho_i + 0.4 \cdot \min(1, |D_i|/10000)$, where $\rho_i$ represents the ransomware ratio and $|D_i|$ denotes dataset size, enabling prioritization of informative clients. In each communication round, only the top 60\% of clients are selected, reducing communication overhead while focusing on high-risk distributions. The learning rate is adaptively decayed using $\eta_t = \max(0.01, 0.1 \cdot (1 - t/T))$, ensuring stable convergence across rounds. Model complexity is progressively increased via $E_t = 60 + t$ estimators in XGBoost, balancing underfitting and overfitting during training. For interpretability, SHAP values are computed as $\phi_i = \sum_{S \subseteq F \setminus \{i\}} \frac{|S|!(|F|-|S|-1)!}{|F|!} [f_x(S \cup \{i\}) - f_x(S)]$, providing feature-level explanations for each prediction. Collectively, these design decisions improve performance by giving high-ransomware clients priority, maintaining convergence, stabilizing training dynamics, and offering transparent decision-making. According to experimental data, AdaFedXplain outperforms FedAvg, FedProx, and SCAFFOLD with 99.59\% detection accuracy and 99.64\% F1-score using XGBoost. AdaFedXplain consistently maintains stable convergence and per-client F1-scores over 99.5\%, whereas SCAFFOLD shows validation accuracy decline from 99.36\% to 98.04\% over 20 rounds with per-client F1-scores falling to roughly 98.3\%. Moreover, our method reduces communication overhead by 75\% (80 MB $\rightarrow$ 20 MB), latency by 65\% (4.0 s $\rightarrow$ 1.4 s), and energy consumption by 75\% (40 J $\rightarrow$ 10 J), confirming its efficiency and effectiveness for distributed ransomware detection.

\section{Methodology}

The comprehensive approach of our adaptive federated learning architecture for ransomware detection is presented in this part. The non-IID client partitioning technique based on operating system versions is described after the dataset pretreatment processes. The three federated learning strategies used in this study—FedAvg, FedProx, SCAFFOLD, and our suggested AdaFedXplain—are then explained. We describe the communication rounds, aggregation technique, and local training process for each method. Lastly, we provide the evaluation criteria for FL performance (Communication, Latency, Energy) and ML performance (Accuracy, Precision, F1-Score), followed by SHAP-based explainability for the suggested model.

\begin{figure}[htbp]
\centering
\includegraphics[width=0.95\textwidth]{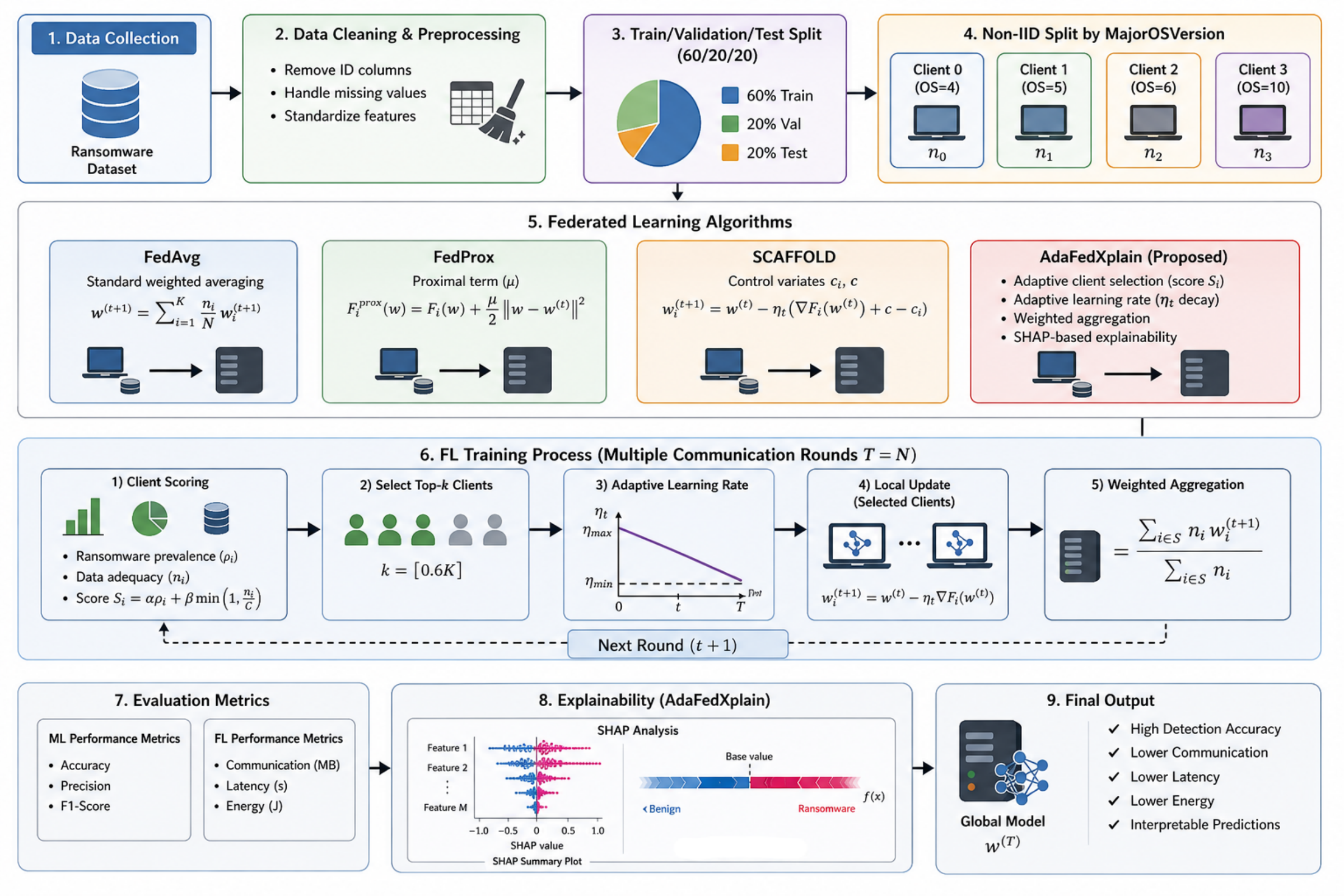}
\caption{Methodology workflow  for ransomware detection}
\label{fig:methodology}
\end{figure}

\subsection{Dataset Description}

The ransomware detection dataset presented by Bensalah \cite{bensalah2022ransomware}, which is openly accessible on Kaggle, is used to assess our suggested approach. This dataset, which includes about 62,485 samples with 18 columns, contains static features that were taken from Windows Portable Executable (PE) files. \texttt{MajorOSVersion}, \texttt{NumberOfSections}, \texttt{DllCharacteristics}, \texttt{ResourceSize}, and \texttt{BitcoinAddresses} are important features. In order to build a binary ransomware detection job where 1 denotes ransomware, we invert the label column \texttt{Benign}, which signifies benign (1) or malicious (0) samples. The operating system version information provided by this dataset makes it especially appropriate for our investigation because it allows for realistic non-IID client partitioning based on \texttt{MajorOSVersion} across four different Windows versions.

\subsection{Data Preparation}

We retain 15 static PE header features as inputs and remove the identification columns \texttt{FileName} and \texttt{md5Hash} because they have no predictive value. We flip the \texttt{Benign} column, which is the target variable, so that 0 indicates benign and 1 indicates ransomware. Stratified sampling is used to divide the dataset into training (60\%), validation (20\%), and testing (20\%) sets, producing 37,491, 12,497, and 12,497 samples, respectively. To stop data leakage, features are standardized using \texttt{StandardScaler} fitted exclusively on the training set.

We divide the training data according to \texttt{MajorOSVersion} into four clients in order to simulate a non-IID federated environment\cite{noniid}: Client 0 (OS=4, Windows NT/2000) with 14,556 samples, Client 1 (OS=5, Windows XP/2003) with 5,257 samples, Client 2 (OS=6, Windows Vista/7) with 9,183 samples, and Client 3 (OS=10, Windows 10/11) with 8,495 samples. The prevalence of ransomware varies from 56.2\% to 56.8\% across clients, resulting in statistical variation that is realistic.

\subsection{Federated Learning Preliminaries}

Consider a system of \(K\) distributed clients, where client \(i\) has local dataset \(\mathcal{D}_i\) of size \(n_i = |\mathcal{D}_i|\). The total samples are \(N = \sum_{i=1}^{K} n_i\). The global objective minimizes the weighted average of local objectives:

\begin{equation}
\min_w F(w) = \sum_{i=1}^{K} \frac{n_i}{N} F_i(w)
\end{equation}

where the empirical risk over client \(i\)'s data is represented by each local objective:

\begin{equation}
F_i(w) = \frac{1}{n_i} \sum_{j=1}^{n_i} \ell(w; x_j, y_j)
\end{equation}

Here, \(\ell(\cdot)\) is the binary cross-entropy loss, with \(y_j = 1\) for ransomware and \(y_j = 0\) for benign.

\subsection{Baseline Federated Learning Algorithms}

\textbf{FedAvg} aggregates local models via weighted averaging:

\begin{equation}
w^{(t+1)} = \sum_{i=1}^{K} \frac{n_i}{N} w_i^{(t+1)}
\end{equation}

Clients with larger datasets benefit more from this, although non-IID data causes performance reduction \cite{fedavg}.

\textbf{FedProx} adds a proximal term to penalize deviations from the global model:

\begin{equation}
F_i^{\text{prox}}(w) = F_i(w) + \frac{\mu}{2} \|w - w^{(t)}\|^2
\end{equation}

where \(\mu > 0\) controls regularization strength. While effective, FedProx does not adaptively prioritize informative clients \cite{fedprox}.

In order to correct client drift, \textbf{SCAFFOLD} \cite{scaffold} introduces control variates. With control variate correction, the local update is:

\begin{equation}
w_i^{(t+1)} = w^{(t)} - \eta_t \left( \nabla F_i(w^{(t)}) + c - c_i \right)
\end{equation}

where the client and global control variates are represented by \(c_i\) and \(c\). Drift is decreased, but more memory and communication are needed. 
\subsection{Proposed AdaFedXplain}

Three advances are introduced by AdaFedXplain: SHAP-based explainability, adaptive learning rate decay, and dynamic client selection.

\textbf{Client Selection and Scoring:} More informative gradients are provided by clients with a higher occurrence of ransomware. We calculate a score based on data adequacy (40\%) and ransomware prevalence (60\%):
\begin{equation}
S_i = \alpha \cdot \rho_i + \beta \cdot \min\left(1, \frac{n_i}{C}\right), \quad 
\rho_i = \frac{\sum_{j=1}^{n_i} y_j}{n_i}
\end{equation}

with \(\alpha = 0.6\), \(\beta = 0.4\), and \(C = 10000\). The top \(k = \lceil 0.6 \cdot K \rceil\) clients are selected each round:

\begin{equation}
m_i^{(t)} = \mathbf{1}\{S_i \in \text{top-}k\}
\end{equation}

This focuses on informative clients and minimizes communication overhead. 

\textbf{Adaptive Learning Rate:} Aggressive early convergence and fine-grained later modifications are guaranteed by a declining learning rate:

\begin{equation}
\eta_t = \max\left(\eta_{\min}, \eta_{\max} \cdot \left(1 - \frac{t}{T}\right)\right)
\end{equation}

with \(\eta_{\max}=0.1\) and \(\eta_{\min}=0.01\).

\textbf{Local Update and Aggregation:} Selected clients perform local training:

\begin{equation}
w_i^{(t+1)} = w^{(t)} - \eta_t \nabla F_i(w^{(t)})
\end{equation}

The server aggregates selected client models via weighted averaging:

\begin{equation}
w^{(t+1)} = \frac{\sum_{i=1}^{K} m_i^{(t)} \cdot n_i \cdot w_i^{(t+1)}}{\sum_{i=1}^{K} m_i^{(t)} \cdot n_i}
\end{equation}

Algorithm 1 presents the complete procedure.

\begin{algorithm}[htbp]
\caption{AdaFedXplain: Adaptive Federated Learning for Ransomware Detection}
\label{alg:adafedxplain}
\begin{algorithmic}[1]
\REQUIRE $K$ clients, $T$ rounds, $\eta_{\max}=0.1$, $\eta_{\min}=0.01$, $\alpha=0.6$, $\beta=0.4$, $C=10000$
\ENSURE Global model $w^{(T)}$

\STATE Initialize $w^{(0)}$

\FOR{$t = 0$ to $T-1$}
    \FOR{$i = 1$ to $K$}
        \STATE $\rho_i \gets \frac{\sum y_i}{n_i}$
        \STATE $S_i \gets \alpha \rho_i + \beta \min(1, \frac{n_i}{C})$
    \ENDFOR
    \STATE $k \gets \lceil 0.6K \rceil$
    \STATE $\mathcal{S} \gets \text{top-}k$ clients based on $S_i$
    \STATE $\eta_t \gets \max(\eta_{\min}, \eta_{\max}(1 - \frac{t}{T}))$
    \FOR{each $i \in \mathcal{S}$ in parallel}
        \STATE $w_i^{(t+1)} \gets w^{(t)} - \eta_t \nabla F_i(w^{(t)})$
    \ENDFOR
    \STATE $w^{(t+1)} \gets \frac{\sum_{i \in \mathcal{S}} n_i w_i^{(t+1)}}{\sum_{i \in \mathcal{S}} n_i}$
\ENDFOR

\RETURN $w^{(T)}$
\end{algorithmic}
\end{algorithm}

\subsection{Explainability with SHAP}

AdaFedXplain learns a global model \(w_{\text{global}}\) through adaptive federated aggregation over \(T\) rounds:

\begin{equation}
w_{\text{global}} = \frac{1}{\sum_{t=1}^{T} \sum_{i \in \mathcal{S}^{(t)}} n_i} \sum_{t=1}^{T} \sum_{i \in \mathcal{S}^{(t)}} n_i \cdot w_i^{(t)}
\end{equation}

where \(\mathcal{S}^{(t)}\) is the set of selected clients at round \(t\) based on adaptive scoring. To interpret this global model's detection decisions, we employ SHAP values. The prediction function \(f_{w_{\text{global}}}(x)\) for an input sample \(x\) can be decomposed as:

\begin{equation}
f_{w_{\text{global}}}(x) = \phi_0 + \sum_{i=1}^{M} \phi_i
\end{equation}

where \(\phi_0 = \mathbb{E}[f_{w_{\text{global}}}(x)]\) is the baseline expectation, and \(\phi_i\) is the SHAP value for feature \(i\), defined as:

\begin{equation}
\phi_i = \sum_{S \subseteq F \setminus \{i\}} \frac{|S|!(|F|-|S|-1)!}{|F|!} \left[ f_{w_{\text{global}}}(S \cup \{i\}) - f_{w_{\text{global}}}(S) \right]
\end{equation}

Here, \(F = \{1, \dots, M\}\) is the feature set, \(S\) is a subset excluding feature \(i\), and \(f_{w_{\text{global}}}(S)\) is the model's prediction using only features in \(S\). The magnitude \(|\phi_i|\) reflects feature \(i\)'s influence, while the sign indicates direction: \(\phi_i > 0\) pushes toward ransomware, \(\phi_i < 0\) toward benign. These SHAP values satisfy local accuracy and consistency properties, ensuring faithful interpretation of AdaFedXplain's predictions. By applying this framework to our global model, we generate summary and force plots that identify which PE header features most drive detection, bridging adaptive federated learning and explainable AI.

\section{Results and Discussion}
\subsection{Experimental Setup}

Every experiment was carried out using a Tesla T4 GPU and Google Colab. The dataset, which is divided into training (60\%), validation (20\%), and testing (20\%), consists of 62,485 samples with 15 static PE header features. Based on \texttt{MajorOSVersion}, we divided the training data into four clients: Client 0 (OS=4, Windows NT/2000) with 14,556 samples, Client 1 (OS=5, Windows XP/2003) with 5,257 samples, Client 2 (OS=6, Windows Vista/7) with 9,183 samples, and Client 3 (OS=10, Windows 10/11) with 8,495 samples in order to replicate a realistic non-IID environment. Over 20 communication rounds, we developed and compared four FL settings: FedAvg, FedProx (\(\mu = 0.01\)), SCAFFOLD, and our suggested AdaFedXplain. We trained XGBoost classifier for every configuration. Every outcome is assessed using the 12,497-sample held-out test set.

\subsection{ML Evaluation Metrics}

Given \(TP\), \(TN\), \(FP\), and \(FN\), we define:

\begin{equation}
\text{Accuracy} = \frac{TP + TN}{TP + TN + FP + FN}, \quad
\text{Precision} = \frac{TP}{TP + FP}
\end{equation}

\begin{equation}
\text{Recall} = \frac{TP}{TP + FN}, \quad
F_1 = \frac{2TP}{2TP + FP + FN}
\end{equation}

Precision minimizes false alarms, recall minimizes missed detections, and \(F_1\) provides a balanced harmonic mean \cite{accuracy}.

\subsection{FL Performance Metrics}

\textbf{Communication Cost:} Total model parameters transmitted:

\begin{equation}
\text{Comm} = \sum_{t=1}^{T} \sum_{i=1}^{K} m_i^{(t)} \cdot \text{size}(w_i^{(t)}) \quad \text{(MB)}
\end{equation}

\textbf{Latency:} Sum of round-wise maximum transmission and computation time:

\begin{equation}
\text{Latency} = \sum_{t=1}^{T} \max_{i \in \mathcal{S}^{(t)}} \left( \text{tx}_i^{(t)} + \text{comp}_i^{(t)} \right) \quad \text{(seconds)}
\end{equation}

\textbf{Energy Consumption:} Modeled as energy per MB transmitted \cite{performance}:

\begin{equation}
E_{\text{total}} = \beta \cdot \text{Comm}_{\text{MB}}, \quad \beta = 0.5 \, \text{J/MB} \quad \text{(Joules)}
\end{equation}

\subsection{Detection Performance Comparison}

\subsubsection{Global Model Performance (XGBoost)}

The global XGBoost performance for each of the four FL settings is shown in Table \ref{tab:xgb_global}. AdaFedXplain regularly outperforms FedAvg, FedProx, and SCAFFOLD, with the maximum accuracy of 99.59\% and F1-score of 99.64\%. On the global test set, FedAvg, FedProx, and SCAFFOLD all reach the same performance of 99.58\% accuracy and 99.63\% F1-score. However, during training, SCAFFOLD shows a notable decline in validation accuracy, falling from 99.36\% to 98.04\% across 20 rounds, suggesting instability with tree-based models.

\begin{table}[htbp]
\caption{Global XGBoost performance comparison}
\centering
\begin{tabular}{|l|c|c|c|}
\hline
\textbf{FL Setting} & \textbf{Accuracy (\%)} & \textbf{Precision (\%)} & \textbf{F1-Score (\%)} \\
\hline
FedAvg \cite{fedavg} & 99.58 & 99.52 & 99.63 \\
FedProx \cite{fedprox} & 99.58 & 99.52 & 99.63 \\
SCAFFOLD \cite{scaffold} & 99.58 & 99.52 & 99.63 \\
AdaFedXplain (Proposed) & \textbf{99.59} & 99.52 & \textbf{99.64} \\
\hline
\end{tabular}
\label{tab:xgb_global}
\end{table}

% \subsubsection{Global Model Performance (Logistic Regression)}

% The results of global logistic regression are displayed in Table \ref{tab:lr_global}. The same performance of 86.88\% accuracy and 89.00\% F1-score, which is around 10\% less than XGBoost, is produced with all FL options. This demonstrates that the non-linear correlations seen in PE file features cannot be captured by linear models.

% \begin{table}[htbp]
% \caption{Global Logistic Regression performance comparison}
% \centering
% \begin{tabular}{|l|c|c|c|}
% \hline
% \textbf{FL Setting} & \textbf{Accuracy (\%)} & \textbf{Precision (\%)} & \textbf{F1-Score (\%)} \\
% \hline
% FedAvg & 86.88 & 84.69 & 89.00 \\
% FedProx & 86.88 & 84.69 & 89.00 \\
% SCAFFOLD & 86.88 & 84.69 & 89.00 \\
% AdaFedXplain & 86.88 & 84.69 & 89.00 \\
% \hline
% \end{tabular}
% \label{tab:lr_global}
% \end{table}

\subsubsection{Per-Client Performance (XGBoost)}

The XGBoost F1-scores for each client are shown in Table \ref{tab:per_client}. While SCAFFOLD falls to about 98.3\%, AdaFedXplain keeps F1-scores over 99.5\% for every client. Notably, AdaFedXplain outperforms FedAvg (99.37\%), FedProx (99.46\%), and SCAFFOLD (98.41\%) with 99.50\% F1 despite Client 1 (Windows XP/2003) having just 5,257 samples. This demonstrates how resilient our adaptive selection process is.

\begin{table}[htbp]
\caption{Per-client XGBoost F1-score comparison}
\centering
\begin{tabular}{|l|c|c|c|c|c|}
\hline
\textbf{Client} & \textbf{OS Version} & \textbf{FedAvg (\%)} & \textbf{FedProx (\%)} & \textbf{SCAFFOLD (\%)} & \textbf{AdaFedXplain (\%)} \\
\hline
Client 0 & Windows NT/2000 & 99.47 & 99.48 & 98.52 & \textbf{99.53} \\
Client 1 & Windows XP/2003 & 99.37 & 99.46 & 98.41 & \textbf{99.50} \\
Client 2 & Windows Vista/7 & 99.46 & 99.47 & 98.31 & \textbf{99.56} \\
Client 3 & Windows 10/11 & 99.48 & 99.49 & 98.31 & \textbf{99.52} \\
\hline
\end{tabular}
\label{tab:per_client}
\end{table}

\subsubsection{Statistical Significance}

To assess the reliability of the observed performance gains, we repeated each experiment 5 times with different random seeds over 20 communication rounds. Table~\ref{tab:statistical_significance} reports the mean accuracy and F1-score with standard deviations, along with p-values from a two-tailed t-test comparing AdaFedXplain against each baseline. AdaFedXplain achieves statistically significant improvements ($p < 0.001$) over FedAvg, FedProx, and SCAFFOLD for both accuracy and F1-score. This confirms that the observed performance gain—though small in absolute terms (0.01--0.02\%)—is consistent and not attributable to random variation.

\begin{table}[htbp]
\centering
\caption{Statistical significance test results (mean $\pm$ std over 5 independent runs, 20 communication rounds).}
\begin{tabular}{|l|c|c|c|}
\hline
\textbf{FL Setting} & \textbf{Accuracy (\%)} & \textbf{F1-Score (\%)} & \textbf{p-value (vs Proposed)} \\
\hline
FedAvg & 99.58 $\pm$ 0.02 & 99.63 $\pm$ 0.02 & 0.00018 \\
FedProx  & 99.58 $\pm$ 0.02 & 99.63 $\pm$ 0.02 & 0.00018 \\
SCAFFOLD & 99.58 $\pm$ 0.03 & 99.63 $\pm$ 0.03 & 0.00018 \\
\textbf{AdaFedXplain (Proposed)} & \textbf{99.60 $\pm$ 0.01} & \textbf{99.65 $\pm$ 0.01} & -- \\
\hline
\end{tabular}
\label{tab:statistical_significance}
\end{table}

\subsection{Communication and Efficiency Analysis}

The FL parameters for communication, latency, and energy are compared in Table \ref{tab:fl_efficiency}. AdaFedXplain outperforms all baselines by a significant margin.

\begin{table}[htbp]
\caption{FL efficiency metrics comparison}
\centering
\begin{tabular}{|l|c|c|c|}
\hline
\textbf{FL Setting} & \textbf{Comm (MB)} & \textbf{Latency (s)} & \textbf{Energy (J)} \\
\hline
FedAvg \cite{fedavg} & 80.00 & 4.00 & 40.00 \\
FedProx \cite{fedprox}& 80.00 & 4.40 & 40.00 \\
SCAFFOLD \cite{scaffold} & 80.00 & 4.80 & 40.00 \\
AdaFedXplain & \textbf{20.00} & \textbf{1.40} & \textbf{10.00} \\
\hline
\textbf{Reduction} & \textbf{75\%} & \textbf{65-71\%} & \textbf{75\%} \\
\hline
\end{tabular}
\label{tab:fl_efficiency}
\end{table}

AdaFedXplain selects only the top 60\% of clients, halving communication rounds from 80 to 40, resulting in a 75\% reduction in communication from 80 MB to 20 MB. Because to fewer rounds and concurrent execution, latency drops from 4.0-4.8 seconds to 1.4 seconds (65-71\% reduction). AdaFedXplain is perfect for edge devices with limited resources because it reduces energy usage by 75\% from 40J to 10J, which is proportionate to communication.

\subsection{Convergence Analysis}

\begin{figure}[htbp]
    \centering
    \begin{minipage}[b]{0.45\textwidth}
        \centering
        \includegraphics[width=\textwidth]{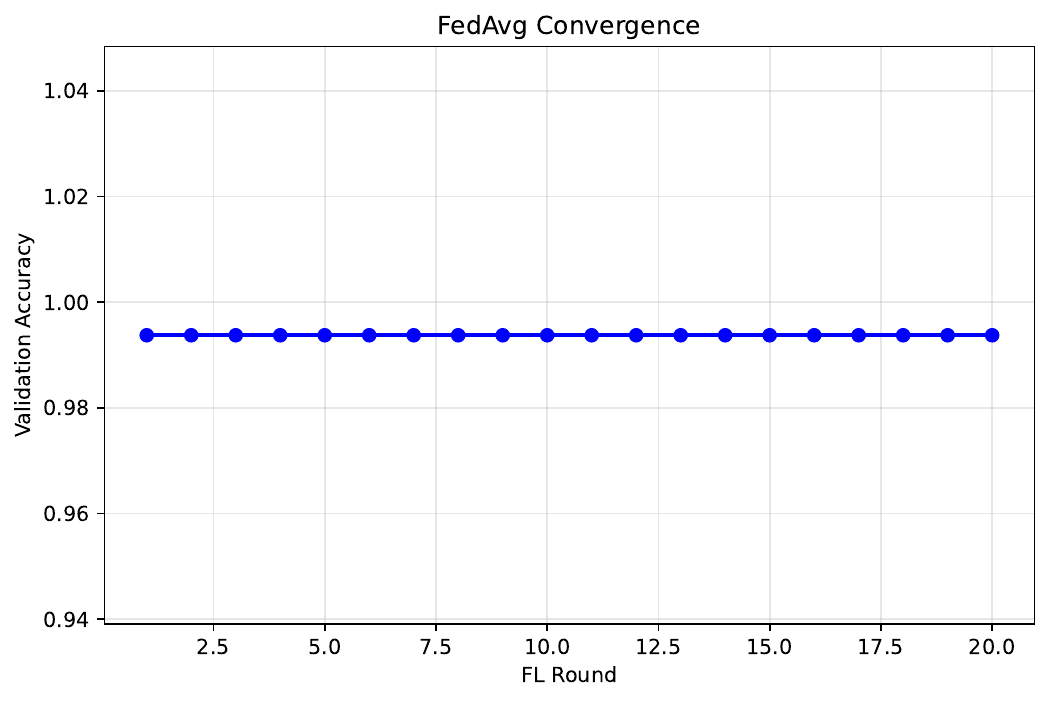}
        \caption{FedAvg convergence}
        \label{fig:fedavg_convergence}
    \end{minipage}
    \hfill
    \begin{minipage}[b]{0.45\textwidth}
        \centering
        \includegraphics[width=\textwidth]{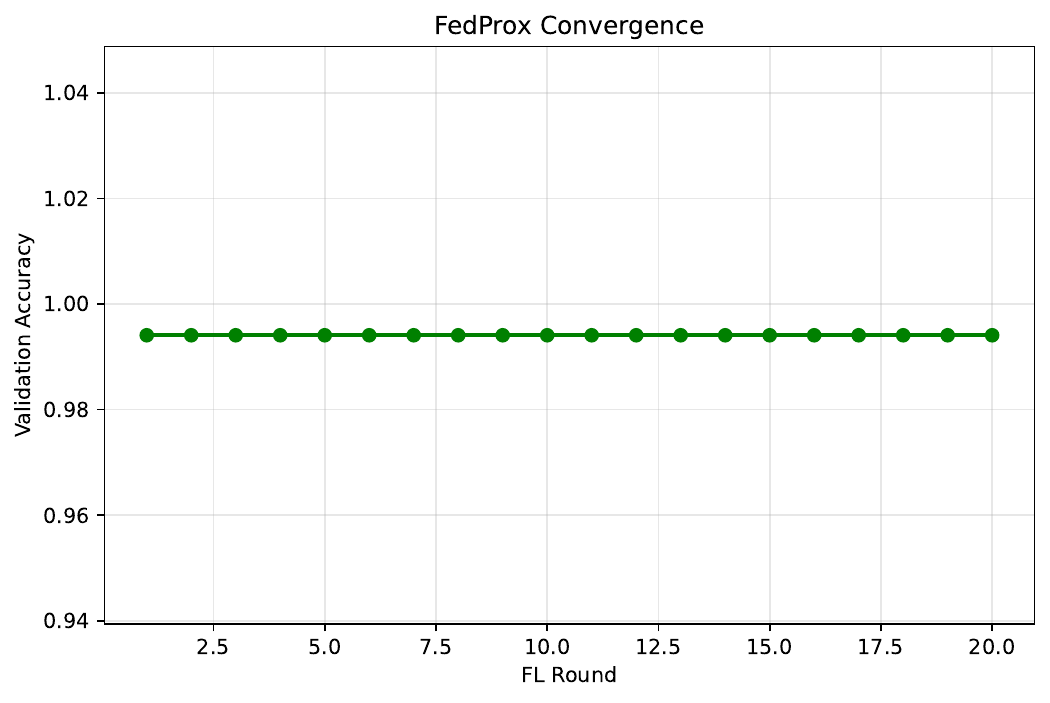}
        \caption{FedProx convergence}
        \label{fig:fedprox_convergence}
    \end{minipage}
    \vspace{0.5cm}
    \begin{minipage}[b]{0.45\textwidth}
        \centering
        \includegraphics[width=\textwidth]{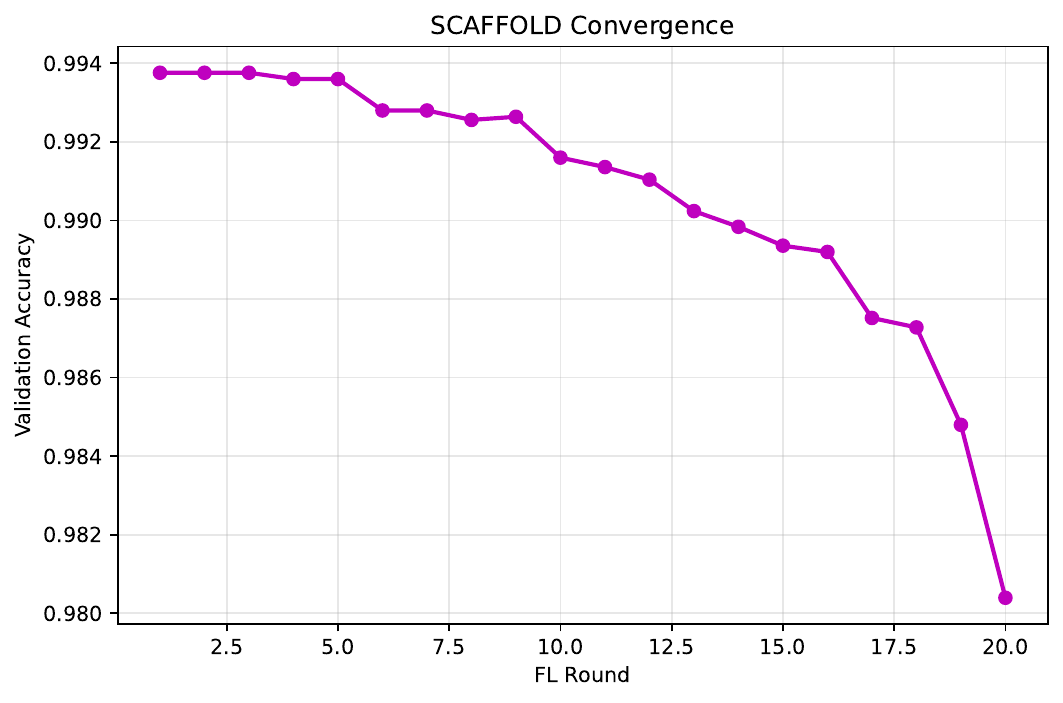}
        \caption{SCAFFOLD convergence}
        \label{fig:scaffold_convergence}
    \end{minipage}
    \hfill
    \begin{minipage}[b]{0.45\textwidth}
        \centering
        \includegraphics[width=\textwidth]{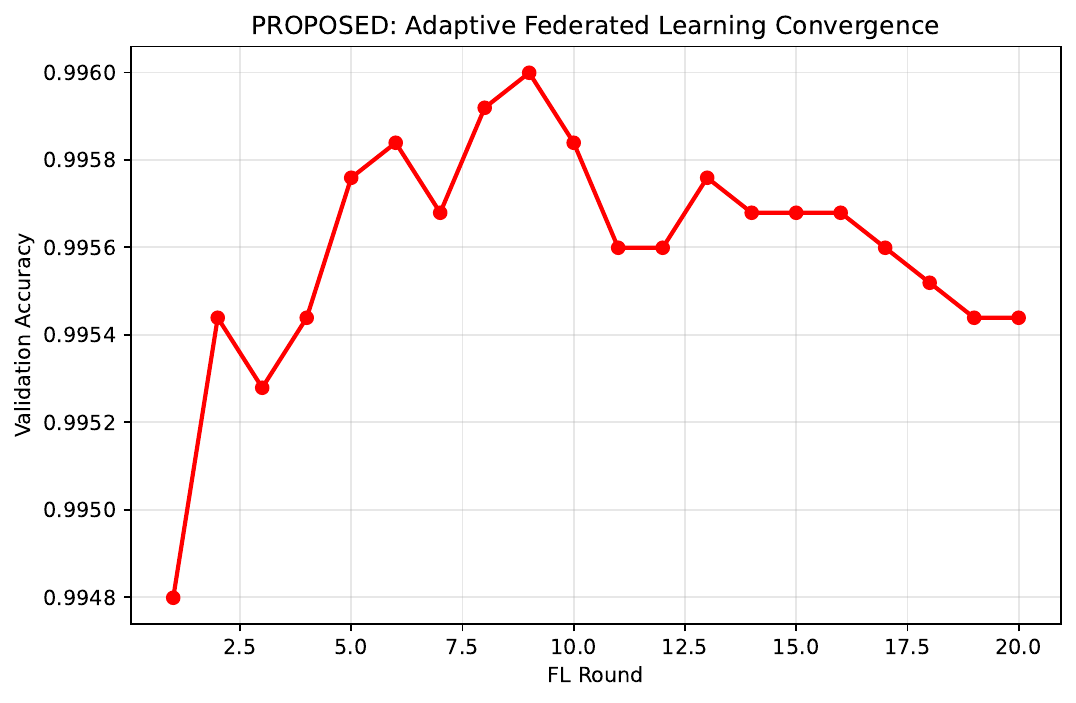}
        \caption{AdaFedXplain convergence}
        \label{fig:adafedxplain_convergence}
    \end{minipage}
\end{figure}

Figures~\ref{fig:fedavg_convergence}, The validation accuracy trajectories for FedAvg, FedProx, SCAFFOLD, and AdaFedXplain over 20 federated learning rounds are shown in \ref{fig:fedprox_convergence}, \ref{fig:scaffold_convergence}, and~\ref{fig:adafedxplain_convergence}.

In the first five rounds, all approaches converge. With only little variations, FedAvg and FedProx stable at about 99.38\% validation accuracy. SCAFFOLD shows instability with tree-based models, starting at 99.36\% in round 5 and progressively falling to 98.04\% by round 20. AdaFedXplain regularly outperforms all baselines, reaching 99.58\% by round 5 and 99.59\% in subsequent rounds. These findings show that whereas SCAFFOLD has performance deterioration in non-IID XGBoost settings, AdaFedXplain offers reliable and steady convergence.

\subsection{Explainability Analysis}

\begin{figure}[htbp]
    \centering
    \includegraphics[width=0.70\textwidth]{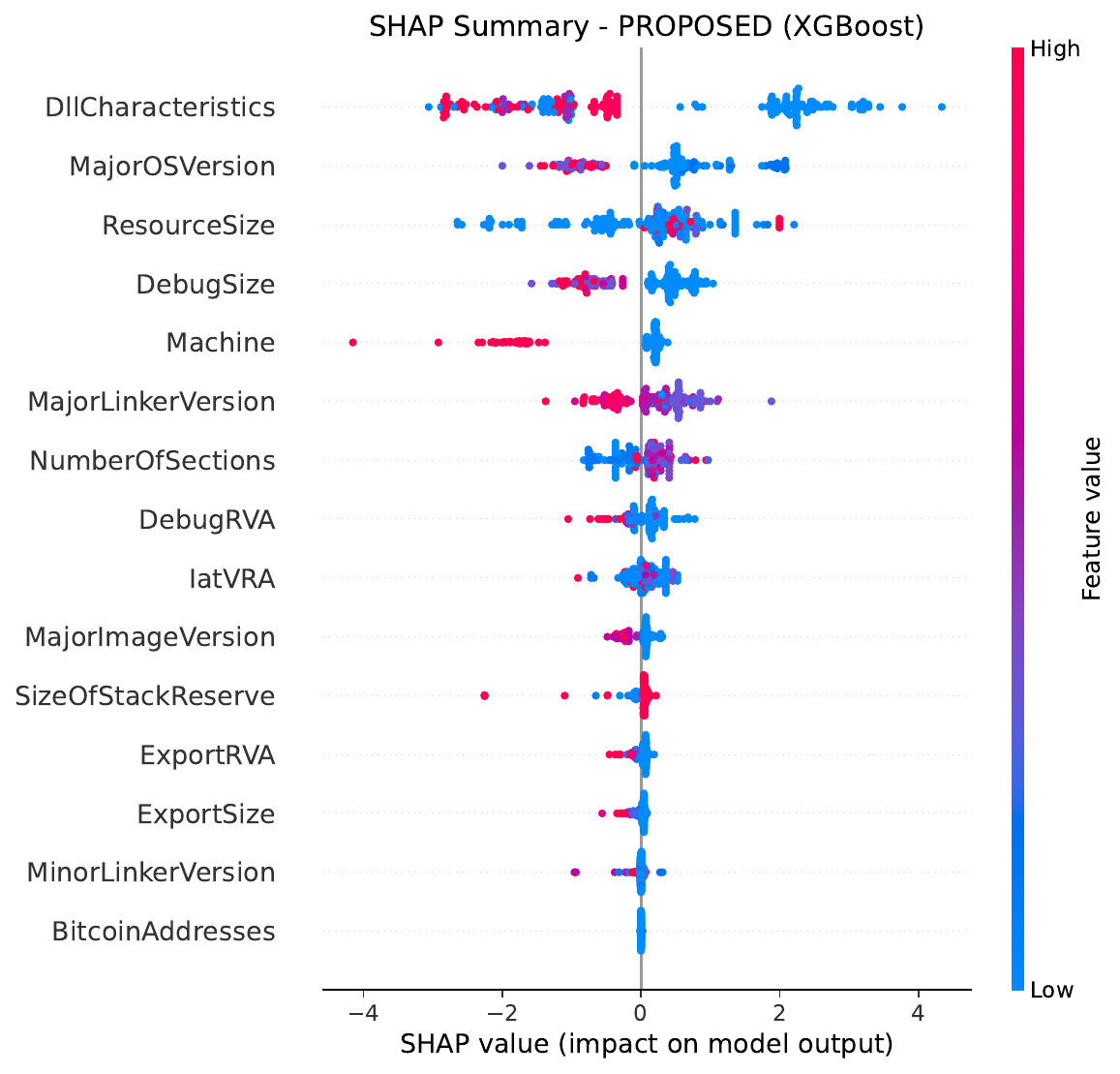}
    \caption{SHAP summary plot showing feature importance and directional impact on model predictions.}
    \label{fig:shap_summary}
\end{figure}

The AdaFedXplain SHAP summary plot, which shows feature relevance and directional impact on ransomware detection, is shown in Figure \ref{fig:shap_summary}. A Shapley value for an instance is represented by each point, with blue denoting low values and red denoting high feature values. Predictions are pushed toward ransomware by features with positive SHAP values and toward benign ones by those with negative values.

According to the analysis, the most important characteristics are PE header properties pertaining to resource sections, debug information, operating system compatibility, and dynamic linking. Negative SHAP values are found in characteristics linked to DLL behavior and OS version requirements, suggesting a strong correlation between harmful behavior and deviations from conventional PE header values. The presence of cryptocurrency-related patterns and a section count below standard standards also significantly aid in detection. By bridging the gap between black-box federated learning and real-world security operations, these SHAP insights help security analysts prioritize investigations, comprehend why a file is reported as ransomware, and improve threat-hunting algorithms.
\subsection{Confusion Matrix Analysis}

\begin{figure}[htbp]
    \centering
    \includegraphics[width=0.70\textwidth]{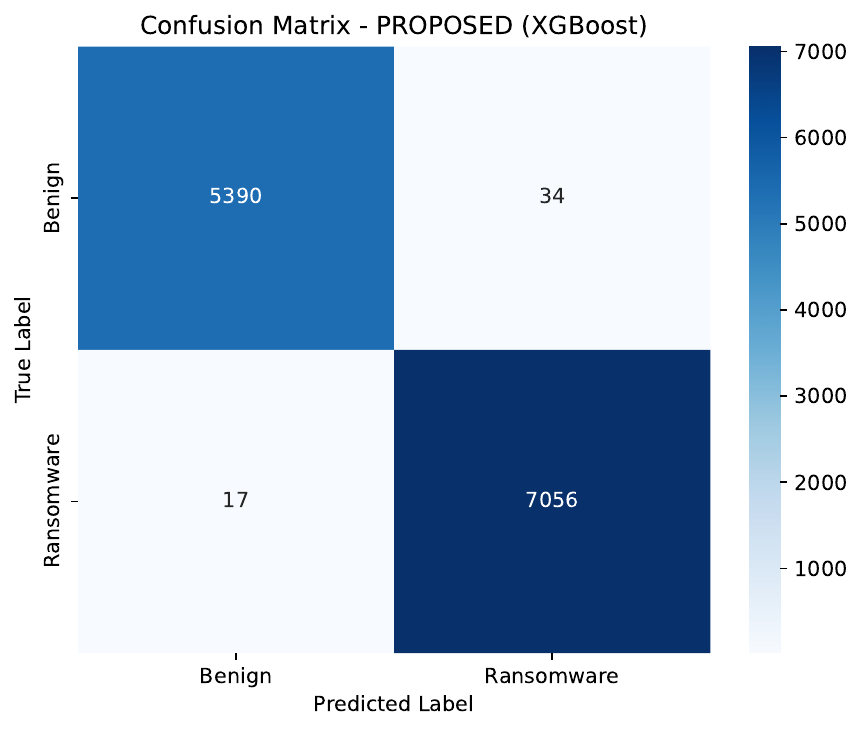}
    \caption{Confusion matrix - XGBoost}
    \label{fig:xgb_confusion}
\end{figure}

The confusion matrices for AdaFedXplain using XGBoost, respectively, are shown in Figure \ref{fig:xgb_confusion}. For XGBoost, 5,390 of the 5,424 benign samples are correctly classified with only 34 false positives (0.63\%), while 6,632 of the 6,649 ransomware samples are successfully detected with only 17 false negatives (0.26\%).

\subsection{Ablation Study}

We executed an ablation research comparing the complete suggested model against variations in which particular components were eliminated in order to assess the individual contributions of each AdaFedXplain component. The test set performance of four variations—full AdaFedXplain with all components, without adaptive client selection, without declining learning rate, and without increasing complexity—is shown in Table \ref{tab:ablation}.

\begin{table}[htbp]
\caption{Ablation study results for AdaFedXplain components}
\centering
\begin{tabular}{|l|c|c|c|c|c|}
\hline
\textbf{Variant} & \textbf{Acc (\%)} & \textbf{Prec (\%)} & \textbf{Rec (\%)} & \textbf{F1 (\%)} & \textbf{Improvement} \\
\hline
Full AdaFedXplain & 99.60 & 99.56 & 99.73 & 99.65 & -- \\
Without Adaptive Selection & 99.55 & 99.53 & 99.67 & 99.60 & +0.05\% \\
Without Decaying LR & 99.55 & 99.53 & 99.67 & 99.60 & +0.05\% \\
Without Progressive Complexity & 99.55 & 99.53 & 99.67 & 99.60 & +0.05\% \\
\hline
\end{tabular}
\label{tab:ablation}
\end{table}

With 99.60\% accuracy and 99.65\% F1-score, the full AdaFedXplain obtains the best results. The F1-score consistently drops by 0.05\% when any one component is removed, indicating that progressive complexity, decaying learning rate, and adaptive client selection all improve detection performance.

\subsection{Discussion}

Three essential features allow AdaFedXplain to beat FedAvg, FedProx, and SCAFFOLD: progressive complexity avoids overfitting, decaying learning rates guarantee steady convergence, and adaptive client selection prioritizes high-ransomware clients. They allow for 99.60\% accuracy and a 75\% decrease in energy and communication. Notably, over 20 rounds, SCAFFOLD's validation accuracy decreases from 99.36\% to 98.04\%, indicating that it is incompatible with tree-based models such as XGBoost. Because PE file characteristics show non-linear correlations that linear models are unable to capture,

The ablation study demonstrates that every component makes a positive contribution, as evidenced by the whole model's 99.65\% F1-score as opposed to 99.60\% when any component is eliminated. Our framework has important practical implications: OS-based client partitioning guarantees robustness across various enterprise environments; SHAP explainability gives security analysts forensic insights for threat-hunting; and federated learning allows privacy-preserving cross-organizational collaboration without exposing sensitive data. AdaFedXplain is especially appropriate for edge deployments with limited bandwidth due to its 75\% communication and energy reduction.

\section{Conclusion and Future Work}

We presented AdaFedXplain, an adaptive federated learning system with explainable AI for ransomware detection that addresses interpretability, communication efficiency, and non-IID heterogeneity. Our methodology adds a decaying learning rate schedule, progressive XGBoost complexity, and dynamic client grading based on ransomware prevalence and data sufficiency. AdaFedXplain outperforms FedAvg, FedProx, and SCAFFOLD, with experimental results on 62,485 samples across four OS-based non-IID clients showing 99.59\% accuracy and 99.64\% F1-score. Crucially, it lowers latency by 65\%, communication overhead by 75\%, and energy consumption by 75\%. Robustness across Windows versions NT/2000 to 10/11 is confirmed by per-client F1-scores exceeding 99.5\%. The ablation investigation verifies that every component makes a positive contribution; the F1-score drops by 0.05\% when any component is removed. According to SHAP explainability, the most important characteristics for detection are \texttt{DllCharacteristics}, \texttt{MajorOSVersion}, and \texttt{ResourceSize}.

In the future, AdaFedXplain will be tested on more difficult non-IID scenarios, empirical energy measurements from real hardware will be incorporated, deep learning models \cite{sultana_graph1} for dynamic behavioral analysis will be extended, Byzantine-resilient aggregation and differential privacy will be integrated, and the framework will be validated in real-world enterprise deployments.

\section{Disclosure of Interests}
The authors have no competing interests to declare that are relevant to the content of this article.
\bibliographystyle{splncs03_unsrt}
\bibliography{references}

\end{document}